\documentclass[reprint, superscriptaddress, nofootinbib, altaffilsymbol, aps, prc, floatfix, longbibliography]{revtex4-1}
\usepackage[colorlinks=true,pdfstartview=FitV,linkcolor=blue,citecolor=blue,urlcolor=blue,bookmarks=false]{hyperref}
\usepackage[usenames,dvipsnames]{color}
\usepackage[sc]{mathpazo}
\usepackage[below]{placeins}
\usepackage{afterpage}
\usepackage[separate-uncertainty,retain-explicit-plus,per-mode = symbol, detect-weight=true, range-phrase=-, range-units=single, tight-spacing=true]{siunitx}
\usepackage{booktabs}
\usepackage{mathtools}
\usepackage{cuted}
\usepackage{longtable}
\usepackage{multirow}
\usepackage{rotating}
\usepackage[mathlines]{lineno}
\usepackage{graphicx}
\usepackage{paralist}
\usepackage{wrapfig}
\usepackage{appendix}
\usepackage{vruler}
\usepackage{etoolbox}
\usepackage{tikz}
\usepackage{listings}
\usepackage{amsmath}
\usepackage{subfigure}
\usepackage{relsize}
\usepackage{etoolbox}
\usepackage{gensymb}
\usepackage{colortbl}
\modulolinenumbers[5]
\newcolumntype{d}[1]{D{.}{\cdot}{#1}}
\newcommand{\trit}{$^3$H}
\newcommand{\ber}{$^7$Be}
\newcommand{\sod}{$^{22}$Na}
\newcommand{\isotope}[2]{$^{#2}${#1}}

\DeclareMathOperator\erfc{erfc}

\DeclareSIUnit\bqkg{Bq/kg}
\DeclareSIUnit\ppt{pg/g}
\DeclareSIUnit\ppb{ng/g}
\DeclareSIUnit\ppm{\ensuremath{\micro}g/g}
\DeclareSIUnit\gpg{g/g}
\DeclareSIUnit\c{$c$}
\DeclareSIUnit\day{day}
\DeclareSIUnit\week{w}
\DeclareSIUnit\year{yr}
\DeclareSIUnit\standard{std}
\DeclareSIUnit\str{sr}

\begin{document}

\title{Removal of spallation-induced tritium from silicon through diffusion}
\newcommand{\pnnl}{Pacific Northwest National Laboratory, Richland, WA 99352, USA}
\newcommand{\gau}{GAU-Radioanalytical,  University of Southampton, Southampton, SO14 3ZH, UK}
\newcommand{\uwashington}{Center for Experimental Nuclear Physics and Astrophysics, University of Washington, Seattle, Washington 98195, USA}
\newcommand{\uchicago}{Kavli Institute for Cosmological Physics and The Enrico Fermi Institute, The University of Chicago, Chicago, Illinois 60637, USA}
\author{R. Saldanha}\email[Corresponding author: ]{richard.saldanha@pnnl.gov}\affiliation{\pnnl}
\author{D. Reading}\email[Corresponding author: ]{d.reading@southampton.ac.uk}\affiliation{\gau}
\author{P.E. Warwick}\affiliation{\gau}
\author{A.E. Chavarria}\affiliation{\uwashington}
\author{B. Loer}\affiliation{\pnnl}
\author{P. Mitra}\affiliation{\uwashington}
\author{L. Pagani}\affiliation{\pnnl}
\author{P. Privitera}\affiliation{\uchicago}

%\date{}

\begin{abstract}
Tritium, predominantly produced through spallation reactions caused by cosmic ray interactions, is a significant radioactive background for silicon-based rare event detection experiments, such as dark matter searches. We have investigated the feasibility of removing cosmogenic tritium from high-purity silicon intended for use in low-background experiments. We demonstrate that significant tritium removal is possible through diffusion by subjecting silicon to high-temperature ($>$ \SI{400}{\celsius}) baking. Using an analytical model for the de-trapping and diffusion of tritium in silicon, our measurements indicate that cosmogenic tritium diffusion constants are comparable to previous measurements of thermally-introduced tritium, with complete de-trapping and removal achievable above \SI{750}{\celsius}. This approach has the potential to alleviate the stringent constraints of cosmic ray exposure prior to device fabrication and significantly reduce the cosmogenic tritium backgrounds of silicon-based detectors for next-generation rare event searches.
\end{abstract}
%---
%---
\maketitle
\section{Introduction}
Silicon is widely used as the sensitive target for low-background rare event searches.
The extremely low radioactive contamination in high purity silicon \cite{DAMIC:2020wkw}, coupled with the ability to achieve eV-scale energy thresholds, has led silicon-based detectors such as DAMIC \cite{SENSEI:2023rcc}, DAMIC-M \cite{DAMIC-M:2023gxo}, SENSEI \cite{SENSEI:2023zdf}, SuperCDMS \cite{SuperCDMS:2024yiv}, and TESSERACT \cite{Billard:2024zvc} to be used in the direct search for dark matter interactions. One of the primary challenges for these experiments is radioactive decays that can mimic a potential dark mater signal. Of particular concern are radioactive isotopes produced within the active silicon substrate of the devices by interactions of high energy cosmic ray particles.

The dominant cosmogenic background is tritium (\trit) \cite{SuperCDMS:2016wui, DAMIC:2021crr} which has a half-life of 12.3 years and decays through a pure beta decay with an endpoint energy of \SI{18.6}{\keV}. The tritium is primarily produced by spallation of silicon nuclei from interactions with secondary cosmogenic particles. At sea level, the production of tritium is roughly 124 atoms/kg/day, dominated by neutron-induced reactions \cite{Saldanha2020-Si}. This corresponds to a low-energy (\SIrange{0}{5}{\keV}) background rate of roughly $0.002$ events/kg/keV/day for each day of sea-level exposure of the silicon. Following the growing of the high-purity silicon boule (which likely removes any non-silicon contaminants from the crystal \cite{VONAMMON198494}), the silicon needs to be sliced into wafers, fabricated into devices, and packaged, before it can be deployed within an experiment. All these steps typically occur at above-ground facilities which can be separated by several thousands of miles. Current generation experiments such as DAMIC-M and SuperCDMS have gone to great lengths to reduce the cosmic ray exposure of the silicon to less than 60-100 days sea-level equivalent, by storing the silicon underground whenever possible and utilizing custom shielded containers during device production and transport \cite{DAMIC-M:2024ooa, orrell2023dbd}. However, next-generation experiments such as Oscura \cite{Oscura:2022vmi} are targeting significantly lower radioactive background levels, which will require additional mitigation strategies. In this work, we aim to determine the feasibility of removing spallation-induced tritium from high purity crystalline silicon through diffusion at elevated temperatures.

\section{Tritium diffusion in silicon}
Hydrogen is widely used in the semiconductor industry to improve the performance of silicon-based devices. The primary motivation for the incorporation of hydrogen into the silicon lattice is to passivate dangling bonds at interfaces or defects, thereby reducing recombination, though there are several other applications \cite{sopori2001silicon, pearton1987hydrogen, chevallier1988hydrogen}. The widespread practice of introducing hydrogen into silicon has led to several studies of hydrogen movement and bonding within the silicon lattice. While in a perfect silicon lattice hydrogen is expected to be an interstitial impurity with a large diffusion constant, it is now understood that hydrogen interacts with nearly all impurities and defects in the lattice, leading to diffusion characteristics that depend strongly on the quality of the silicon material, the method of hydrogen incorporation, and the molecular and charge state of the hydrogen within the lattice \cite{pearton1987hydrogen, myers1992hydrogen}. For these reasons, measurements of the diffusion constant of hydrogen in silicon vary by several orders of magnitude (see for example Figure 7 in Ref.~\cite{sopori2001silicon}), though in most cases the diffusion constant is found to be $>$ \SI{1E-8}{\cm\squared\per\sec} above \SI{700}{\celsius}.

The relatively large diffusion constant of hydrogen in silicon raises the possibility of removing cosmogenically produced tritium from silicon substrates used for rare event searches \cite{martoff1987limits}. However, previous results in the literature indicate that the diffusion constant depends strongly on the method by which tritium was incorporated into the silicon. Ichimiya and Furuichi \cite{ichimiya1968solubility} thermally diffused tritium into single-crystal silicon at high temperatures and measured diffusion constants roughly consistent with other measurements of hydrogen in silicon. Saeki et. al. \cite{saeki1983origins} looked for differences in the diffusion constant between tritium injected thermally and tritium injected as recoils from nuclear reactions. They found that for the recoil-injected tritium, the diffusion constants were orders of magnitude lower than those for thermally-injected tritium and also had a significantly stronger temperature dependence. The difference in the measured diffusion for the two methods was attributed to chemical trapping and radiation-induced damage and defects in the crystal. Cosmogenic tritium is primarily produced by neutron-induced spallation of the silicon nucleus \cite{Saldanha2020-Si}, a very energetic process that is likely to cause damage to the lattice through both the initial nuclear interaction and the stopping of the energetically ejected triton particle. Thus, if cosmogenic tritium also has significantly retarded diffusion, similar to recoil-induced tritium, removal of the tritium may not be feasible on a reasonable time scale. In this work we aim to directly measure the diffusion constants and removal fractions as a function of temperature.

\section{Experimental Plan}
 We introduced tritium into silicon wafer samples using a high intensity, broad spectrum neutron beam at the Los Alamos Neutron Science Center (LANSCE) and measured the removal using a pyrolyser and liquid scintillation counting at GAU-Radioanalytical laboratory.

The neutron beam at the ICE-HOUSE II facility \cite{lisowski2006alamos, icehouse} has a similar energy spectrum to that of cosmic ray neutrons, but with a flux $\sim5 \times 10^8$ times larger than the natural sea-level flux. This facility is well-suited for cosmogenic activation studies and has previously been used to measure tritium, \ber, and \sod~production rates in silicon \cite{Saldanha2020-Si}. The close similarity of the neutron beam spectrum to that of the natural cosmogenic neutrons at the Earth's surface is critical for this measurement as it allows for the automatic incorporation of particle and energy-dependent secondary effects, such as localized damage of the silicon lattice at the site of the triton production and implantation, that can affect the subsequent diffusion and removal of tritium from the silicon samples. Additionally, the high beam flux allows for the production of easily measurable amounts of tritium in the silicon samples with a beam exposure of just a few hours.

The use of a pyrolyser allows for the extraction and collection of tritium from the wafers with high efficiency over a wide range of temperatures. Additionally, by collecting the tritium in batches, one can measure the time profile of the extracted tritium and calculate diffusion coefficients. GAU-Radioanalytical, based at the University of Southampton UK, is a long-established and experienced laboratory specializing in the extraction and quantification of volatile radionuclides for nuclear and radiopharmaceutical facility decommissioning, land remediation, and environmental monitoring surveys. Operating multiple Pyrolyser 6-Trio systems (Raddec International, UK), the laboratory can process a diverse range of sample types, including problematic / complex sample matrices \cite{warwick2010effective}. Tritium measurement is performed using Quantulus 1220 ultra-low-level liquid scintillation counters (LSC) \cite{schafer2000low}. The laboratory is accredited to ISO/IEC 17025:2017 for the extraction and quantification of tritium.

\section{Tritium Generation} 
\subsection{Silicon Wafers} 
23 high-purity silicon wafers were purchased from Virginia Semiconductor Inc. (Virginia, USA) for this work. The wafers were sliced from a float-zone (FZ) grown single-crystal boule with a resistivity of $>$\SI{10000}{\ohm\cm}, matching the silicon substrate purities typically used in rare-event searches. Each wafer is \SI{76.2 \pm 0.3}{\milli\meter} in diameter, with a thickness of \SI{1000 \pm 25}{\micro\meter} and a mass of roughly \SI{10.6}{\gram}. The wafers are undoped, polished on one side, and have a $\langle$100$\rangle$ crystal-plane alignment.

\subsection{Beam Exposure}
The samples were irradiated at the LANSCE WNR ICE-HOUSE II facility~\cite{icehouse} on Target 4 Flight Path 30 Right (4FP30R). A broad-spectrum (0.2--800 MeV) neutron beam was produced via spallation of 800 MeV protons on a tungsten target. A 2-inch (\SI{50.8}{\milli\meter}) diameter beam collimator was used to restrict the majority of the neutrons to within the diameter of the wafers. The neutron fluence was measured with $^{238}$U foils by an in-beam fission chamber~\cite{wender1993fission} placed downstream of the collimator. The beam has a pulsed time structure, which allows the incident neutron energies to be determined using the time-of-flight technique (TOF)---via a measurement between the proton beam pulse and the fission chamber signals~\cite{lisowski2006alamos,wender1993fission}.

The beam exposure of the wafers took place over two days between November 3$^{\mathrm{rd}}$ and 5$^{\mathrm{th}}$, 2019. All 23 wafers were placed back-to-back in a plastic wafer holder and centered on the beam line with the individual wafers separated by roughly \SIrange{3}{6}{\mm} and the ends of the plastic holder removed to reduce attenuation of the beam. The wafers were numbered by their position on the beam line (1 was furthest upstream and 23 furthest downstream), and the wafer labels were tracked throughout the experiment. The wafers were exposed to the beam from 06:38 on Nov.\,3, to 08:13 on Nov.\,5. Following the exposure the wafers were kept in storage for approximately eleven weeks to allow short-lived radioactivity to decay prior to shipment from LANSCE to Pacific Northwest National Laboratory (PNNL).

\subsection{Neutron Fluence}
The total fluence of neutrons during the silicon wafer beam exposure, as measured by the fission chamber, was \num{9.71 \pm 0.77 E12} neutrons above \SI{10}{\MeV}. The energy spectrum of the neutrons was the same (within statistical uncertainties) as the spectrum shown in Figure 5 of Ref.~\cite{saldanha2023cosmogenic} where NaI crystals were exposed to the LANSCE beam just prior to the silicon wafers. As described in detail in Ref.~\cite{saldanha2023cosmogenic}, the uncertainty in the neutron fluence is dominated by the systematic uncertainity in the $^{238}$U(n, f) fission cross section used to monitor the fluence.
While the nominal beam diameter was set by the 2-inch collimator, the cross-sectional beam profile has significant tails at larger radii. At the fission chamber approximately 13\% of neutrons fall outside a 3-inch diameter, as calculated with the beam profile provided by LANSCE. Additionally the beam is slightly diverging, with an estimated cone opening angle of 0.233\degree. A Geant4 \cite{agostinelli2003geant4,allison2016recent} simulation that included the measured beam profile and beam divergence, the measured neutron spectrum, and the full setup geometry (location and materials of the wafers, holder, mounting apparatus, and fission chamber~\cite{wender1993fission}), was used to calculate the neutron fluence through the wafers. Additional details on the Geant4 simulation can be found in Ref.~\cite{saldanha2023cosmogenic}. Including all geometrical effects, attenuation through the wafers, and systematic uncertainties, we estimate that a total of \num{8.59 \pm 0.68 E12} neutrons above \SI{10}{\MeV} passed through the first silicon wafer (wafer 1) and \num{8.00 \pm 0.64 E12} neutrons above \SI{10}{\MeV} passed through the last wafer (wafer 23).

\subsection{Gamma Counting Verification}
For isotopes with half-lifes that are long compared to the 2-day beam exposure, the predicted isotope activity, $P$ [Bq], produced by the beam is given by
\begin{linenomath*}
\begin{align}
\label{eq:beam_act}
P = \frac{n_a}{\tau} \int S(E) \cdot \sigma(E)~dE
\end{align}
\end{linenomath*}
where $n_a$ is the areal number density of the target silicon atoms [\si{atoms\per \cm\squared}], $\tau$ is the mean life [\si{\second}] of the isotope decay, $S(E)$ is the energy spectrum of neutrons [\si{neutrons \per \MeV}], and $\sigma(E)$ [cm$^2$] is the production cross section. Previous measurements of the production rate of \isotope{Na}{22} on the same LANSCE neutron beam \cite{Saldanha2020-Si} allow us to predict the induced activity in the wafer without knowing the cross section (which has not been experimentally measured across the entire energy range of interest). Scaling the result in Ref.~\cite{Saldanha2020-Si} by the relative change in target thickness and neutron fluence, we predicted the \isotope{Na}{22} activity in wafer 1 and 23 to be roughly  \SI{2.2 \pm 0.}{\becquerel} and \SI{2.1 \pm 0.}{\becquerel} respectively.

Following the beam exposure and cooldown, the two wafers at either end of the beam line (wafers 1 and 23) were gamma counted to determine the beam-induced \isotope{Na}{22} activity and cross-check the neutron beam exposure. The wafers were individually placed in a low-background germanium counter \cite{sharma2021sensitivity} in the Shallow Underground Laboratory \cite{aalseth2012shallow} at PNNL and counted for \SIrange{3}{4}{days} in March 2021. The activity of \isotope{Na}{22} in wafers 1 and 23, decay-corrected to the end of the beam exposure, was found to be \SI{3.00 \pm 0.12}{\becquerel} and \SI{2.85 \pm 0.12}{\becquerel} respectively, where the uncertainty is dominated by the uncertainty in the efficiency calibration. While the measured relative activities of the two wafers is in good agreement with the prediction, the absolute measured rates are roughly 30\% larger than expected.

\subsection{Tritium Production}
Spallation can produce light nuclei (e.g. tritons) with kinetic energies that are a significant fraction of the incoming neutron energy. Due to their small mass, these nuclei have relatively long ranges and can therefore be ejected from their volume of creation and implanted into another volume. The measured activity therefore depends not only on the thickness of the target but also on the nature and geometry of the surrounding materials.

The Geant4 beam simulation described above was also used to keep track of the origin and final location of all triton nuclei generated within the geometry. For each wafer we evaluated the tritium activity that was produced within the wafer, the activity that was ejected into the surroundings, activity that was implanted into the wafer from the surroundings, and finally the residual activity remaining in the wafer (residual = produced - ejected + implanted). After accounting for variation in the expected activity from different Geant4 physics lists,  including the previously determined scaling factors \cite{Saldanha2020-Si}, we estimate the tritium activity in wafers 7 to 23 to be \SI{1.11 \pm 0.17}{\becquerel}. Wafers upstream of wafer 7 had smaller estimated residual activity, due to a smaller contribution of implanted tritium, and were not used for this study.

\section{Tritium Extraction and Measurement}
\subsection{Dicing}
In order to fit the wafers into the Pyrolyser, the wafers were diced into six 0.5"-wide slices at the Center for Experimental Nuclear Physics and Astrophysics (CENPA) at the University of Washington, Seattle, using a CO$_2$ laser and glass backing \cite{chung2006silicon}. To ensure that the dicing did not significantly heat up the wafer and cause some of the tritium to diffuse out, we attached temperature sensors to test wafers and adjusted the laser power and time intervals between cuts such that the wafer temperature remained below \SI{100}{\celsius} and the total duration of dicing for each wafer was less than 15 minutes. Extrapolating our diffusion model (presented later in  Section~\ref{sec:detrap_model}) to lower temperatures, we expect less than 2\% of the total tritium to have been removed during dicing. The diced wafers were then shipped to the University of Southampton.  We note that because the natural cosmogenic activity of tritium in the wafers (\SI{15 \pm 3}{\micro\becquerel} at sea-level saturation \cite{Saldanha2020-Si}) is negligible in comparison to the beam-induced activity, no effort was made to reduce activation during shipping.

\subsection{Pyrolyser}  
\begin{figure}
\centering
\includegraphics[width=1\linewidth]{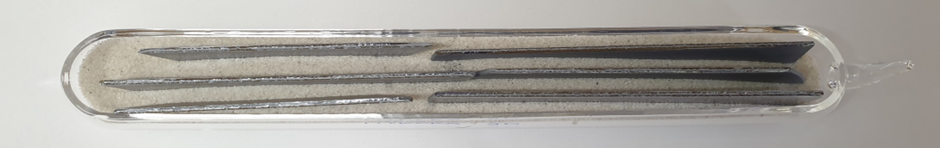}
\caption{Diced wafer loaded in quartz sample boat.}
\label{fig:wafer_boat}
\end{figure}

The time profiles of tritium outgassing from silicon wafers at elevated temperatures were obtained using a Pyrolyser 6-trio furnace, which is a multi-sample, flow-through, combustion/heating system \cite{warwick2010effective}. Briefly, a sample is loaded into a high-purity quartz boat and inserted into a high-purity quartz work tube. The sample is heated/combusted to a target temperature (T $\leq$\SI{1000}{\celsius}) that promotes the off-gassing of volatile radionuclides. A stream of air flows through the work tube and over the sample during heating and carries the combustion gases over a 0.3\% Pt-Al catalyst bed heated at \SI{800}{\celsius}. The catalyst promotes conversion of any tritium to tritiated water. The tritiated water then exits the work tube and is trapped in a bubbler of 0.1M HNO\textsubscript{3}. The work tubes used in this campaign have a nominal internal capacity of 1L and therefore one complete purge of process gases at the typical flow rate of \SI{200}{\milli\litre\per\min} is expected to take approximately 5 minutes.

The diced wafer pieces were placed on a bed of acid washed sand in a quartz sample boat (Figure~\ref{fig:wafer_boat}). Care was taken to prevent the pieces from coming in contact with one another, allowing for maximum surface exposure to the flow gases and to ensure uniform heating. Each wafer desorption experiment was conducted separately, with the Pyrolyser pre-heated to a set target temperature (400, 500, 750, and \SI{1000}{\celsius}) before the wafer was loaded.

\subsubsection{Bubbler change schedule}

\begin{figure}
    \centering
    \includegraphics[width=1\linewidth]{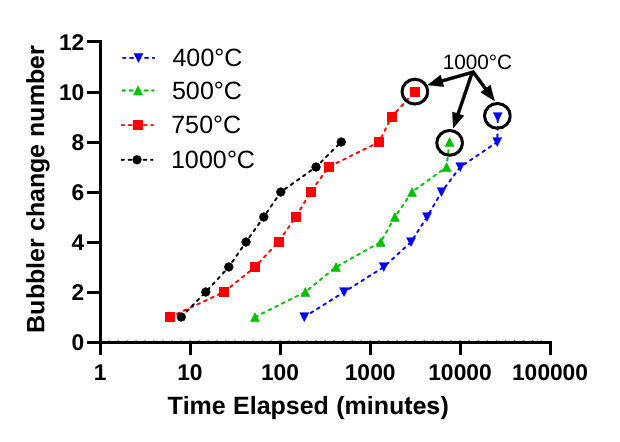}
    \caption{Bubbler change times at each temperature}
    \label{fig:bubbler elapsed timel}
\end{figure}

In order to obtain a time profile of the removed tritium for each wafer,  we periodically changed the Pyrolyser bubblers that collect the tritium. The desired schedule of the bubbler changes was determined by both the expected diffusion coefficient at a given temperature and the minimum detectable activity of the liquid scintillation counting. For planning purposes we assumed diffusion coefficients according to the measurements of Ichimiya and Furuichi \cite{ichimiya1968solubility} and set bubbler changes such that the collected tritium in each bubbler would be roughly three times the minimum detection limit (see Section~\ref{sec:lsc} for more details). The final schedule of bubbler changes was determined by the constraints of the laboratory's working day. The total run time to complete the tritium extraction varied between 8 hours for the measurements at \SI{1000}{\celsius} and 440 hours (18 days) for the measurements at \SI{400}{\celsius} (see Figure~\ref{fig:bubbler elapsed timel}).

In order to determine the fraction of tritium removed at each temperature (independent of our estimates based on the neutron fluence), at the end of the scheduled extraction period for the targeted temperature we ramped up the temperature of the Pyrolyser to \SI{1000}{\celsius} and collected any residual tritium for a minimum of 8 hours. Based on previous measurements in the literature, this is sufficient to remove all residual tritium from the silicon substrate \cite{brodsky1977quantitative, qaim1978triton}.

\subsubsection{Adjustments to Pyrolyser setup} 
The recommended gas flow rate of \SI{200}{\milli\litre\per\min} causes evaporative losses in the bubbler of approximately \SI{300}{\micro\litre\per\hour} at room temperature (\SI{20}{\celsius}). For the typical combustion/heating time of non-organic-rich materials of roughly 6 hours, these losses are relatively small and can be accounted for using tritium recovery factors. However, the desorption profiles for the lower temperature measurements (400 and \SI{500}{\celsius}) require bubblers to be connected to the Pyrolyser for extended periods of time (up to 18 days), in which case the evaporative losses become significant and need to be minimized. Therefore, the bubbler temperature and flow rate for these measurements were reduced and assessed. Reducing the temperature of the bubbler in a water bath to \SI{7}{\celsius} but maintaining a flow rate of \SI{200}{\milli\litre\per\min} reduced the evaporative losses to \SI{113}{\micro\litre\per\hour} (equivalent to 14\% loss per day). Reducing the bubbler temperature further to \SI{0.5}{\celsius} and restricting the flow rate to \SI{50}{\milli\litre\per\min} reduced the evaporative losses to \SI{25}{\micro\litre\per\hour} (equivalent to 0.6\% loss per day). In this configuration (\SI{0.5}{\celsius}, \SI{50}{\milli\litre\per\min} flow rate), the total loss from the bubbler over the duration of the measurements at 400 and \SI{500}{\celsius} was within the measurement and method uncertainties. It should be noted that the reduction in flow rate also increases the purge-time of the work tube but this is insignificant in contrast to the time the bubblers were connected.

\subsection{Liquid scintillation counting}
\label{sec:lsc}
The bubbler solutions (typically \SIrange{8}{9}{\milli\litre} 0.1M HNO\textsubscript{3}) were transferred to a high density polyethylene scintillation vial and were mixed with \SI{12}{\milli\litre} Goldstar scintillation cocktail (Hidex Oy, Finland). Measurements were conducted using a Quantulus 1220 LSC \cite{schafer2000low} and were calibrated using a certified tritium standard (PTB, Germany) where the counting efficiency is determined as a function of quench (the efficiency of light detection through the test sample as determined by an external \isotope{Eu}{152} standard). Samples were dark-adapted for a minimum of 8 hours to reduce the effects of chemiluminescence on the measurement. Samples derived from the \SI{1000}{\celsius} bakes were counted for 8 hours each. All other samples from the remaining bakes were counted for 10 hours each as the activity was expected to be lower. Counting efficiencies for samples were typically 17-18\%. An instrument blank was prepared for each batch of samples and was used to correct for instrument background. Additionally, a quality control standard of known tritium concentration was counted with each batch to verify instrument performance. The limit of detection for an 8 hour count was determined to be 0.0017 Bq/g of wafer.

\section{Results}
\begin{figure*}[t]
    \centering
    \includegraphics[width=6in]{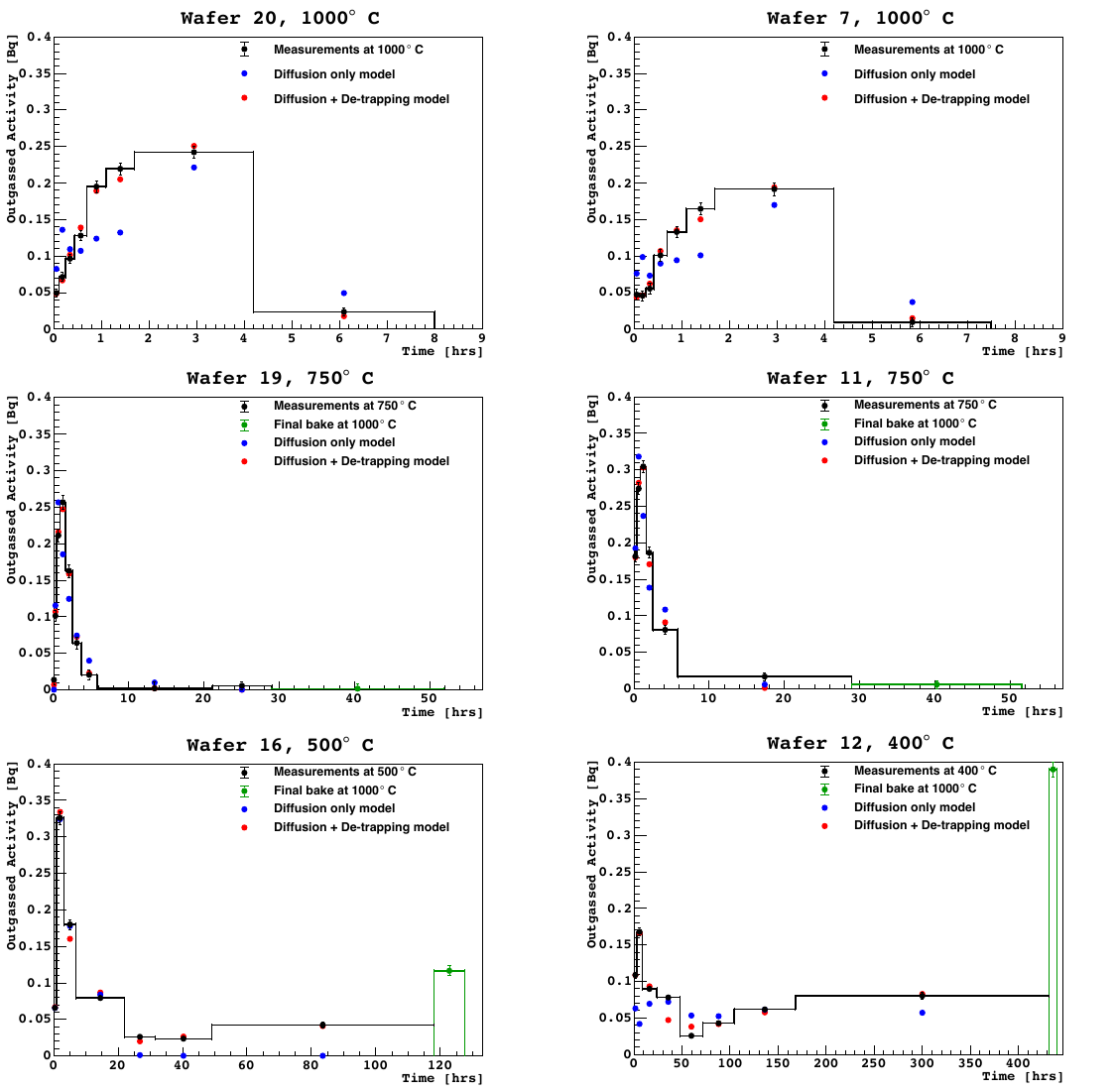}
    \caption{Time profile of tritium outgassing from silicon wafers at different temperatures. The black bin edges mark the start and end of each bubbler change while the central marker shows the measured activity and uncertainty for each bubbler. Results of the final bake at \SI{1000}{\celsius} are shown in green. The blue and red markers show the best-fit result to a diffusion-only and diffusion + de-trapping model (see Section~\ref{sec:fitting} for details).}
    \label{fig:3H evo profiles}
\end{figure*}
Six wafers were randomly chosen for the measurements and the time profiles of tritium removed from each wafer are shown in Figure \ref{fig:3H evo profiles}, including the final bake at \SI{1000}{\celsius} to collect any residual tritium. The measurements were made between May 2021 and December 2022, with all data decay-corrected for the tritium half-life (12.312 years) to the end of the wafer beam-line exposure (5th Nov 2019). 

\subsection {Counting uncertainties \& LSC stability} 

The standard systematic uncertainty was calculated to be 1.2\%. This includes uncertainty contributions associated with sample weighing, the calibration standard, instrument calibration, and decay correction. The systematic uncertainty is then combined with the counting statistical uncertainty to determine the combined standard uncertainty.

All measurements were conducted using the same Quantulus instrument (GAU-Radioanalytical Quantulus no. 6). The long-term stability of the Quantulus instruments has been assessed previously by the laboratory \cite{gaca2017liquid}, and the instrument used in this study was calibrated three times around this study as part of the laboratories annual calibration schedules. As mentioned earlier, the counting efficiency of every batch of samples was checked by measuring an in-house standard of known tritium activity. The measurement of the tritium standard must produce a result within 3$\sigma$ of the control limits established by Shewhart Control Charts for the corresponding wafer measurements to be considered valid. In the event of a failure, the entire sample batch must be recounted. The counting efficiency (\%) for the 8 ml 0.1M HNO\textsubscript{3} + 12 ml Goldstar cocktail was 19.44 ± 0.93, 18.33 ± 0.77 and 19.01 ± 0.67 (uncertainties at coverage factor k=2) with quench values of 711.2, 706.4 and 713.0 for years 2021, 2022 and 2023 respectively.

\subsection{Method accuracy}
For nuclear decommissioning and environmental samples analyzed at GAU-Radioanalytical, the recovery factor of tritium using the Pyrolyser is \SI{90 \pm 12}{\percent}, where the variability is due to sample composition, the nature of tritium binding within the sample, and tritium memory-effect (retention of tritium on the quartz worktube and/or catalyst). Due to the stable wafer composition and the extended time that each wafer was held at the target temperature compared to the purge time, these factors were deemed to be insignificant and no correction was made for collection efficiency.

The combustion and collection of tritium coupled with the measurement by a Quantulus liquid scintillation counter has been assessed by participation in proficiency test exercises (PTE) coordinated by the National Physical Laboratory (NPL, UK) and the Department of Energy (DOE, USA). In the 20 years of participating in the NPL PTE, the average tritium deviation to the reference value is +4.2\% with submitted vs certified values R\textsuperscript{2} = 0.998. The US DOE exercises only had tritium in water samples between 2012-15 where the average deviation to the reference value is +5.1\% with submitted vs certified values R\textsuperscript{2} = 0.981. These small deviations are well within the accepted values for the test exercises and no additional systematic correction is applied to the results. We note that any overall systematic error would only impact the absolute \trit~activity measured and would not impact the time profiles.

\section{Analysis}

\subsection{Tritium fraction removed}
\begin{figure}[t]
    \centering
    \includegraphics[width=\linewidth]{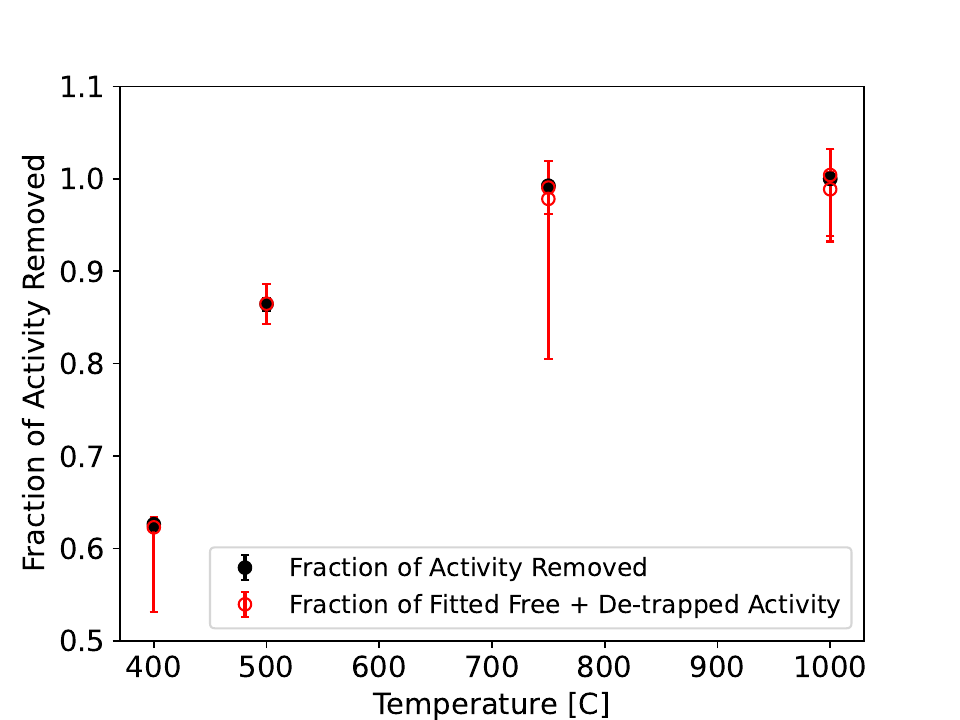}
    \caption{Fraction of tritium removed from the silicon wafers at different baking temperatures. The black data points show the experimentally measured fraction removed compared to the total (data point at \SI{1000}{\celsius} assumes complete removal). The red data points are the results from the fit of the time profile to the diffusion and de-trapping model, including both the free and de-trapped contributions.}
    \label{fig:frac_removed}
\end{figure}

We calculated the fraction of tritium removed by summing the measured activity in each bubbler at a fixed temperature and comparing to the total activity (including the residual activity measured by the final bake at \SI{1000}{\celsius}), as shown in Table~\ref{tab:removal_frac} and Figure~\ref{fig:frac_removed}. Even at the relatively low temperature of \SI{400}{\celsius}, more than half of the neutron-induced tritium can be removed from silicon wafers. The removal fraction increases with temperature as expected, with complete removal at \SI{750}{\celsius}. 
While the total activity measured in some wafers is consistent with the predicted activity from the beam exposure, \SI{1.11 \pm 0.17}{\becquerel}, only \SIrange{0.75}{0.86}{\becquerel} was measured in other wafers. The cause of this variation is unknown. No correlation of the total activity with position of the wafer on the beamline, time between beam exposure and measurement, or baking temperature and gas flow, is found. To further investigate this, the measurements at \SI{750}{\celsius} and \SI{1000}{\celsius} were repeated with two different wafers (11 and 7), however similar variation was observed. Despite the variation in the total activity, the time profile of removed activity (shown in Figure~\ref{fig:cuml_fits}) was very consistent across wafers baked at the same temperature, indicating that the variation in total activity already existed at start of the baking.

\setlength{\tabcolsep}{5pt}
\renewcommand{\arraystretch}{1.5}
\begin{table*}[t]
   \centering
   \begin{tabular}{l|c|c|cc|cc} % Column formatting, @{} suppresses leading/trailing space
       & \SI{400}{\celsius} & \SI{500}{\celsius} & \multicolumn{2}{c|}{\SI{750}{\celsius}} & \multicolumn{2}{c}{\SI{1000}{\celsius}} \\
        \hline
       Wafer & 12 & 16 & 19 & 11 & 20 & 7 \\
       Beam to meas. [days] & 692 & 820 & 769 & 1105 & 549 & 943 \\
       Activity removed [Bq] & \num{0.65 \pm 0.01} & \num{0.74 \pm 0.01} & \num{0.84 \pm 0.02} & \num{1.04 \pm 0.02} & \num{1.03 \pm 0.02} & \num{0.75 \pm 0.02}\\
       Total activity ($A_{tot}$) [Bq] & \num{1.04 \pm 0.01} & \num{0.86 \pm 0.01} & \num{0.84 \pm 0.02} & \num{1.05 \pm 0.02} & \num{1.03 \pm 0.02} & \num{0.75 \pm 0.02}\\
       Residual activity [Bq] & \num{0.39 \pm 0.01} & \num{0.116 \pm 0.007} & $< 0.013$ & $< 0.016$ & - & - \\
       Fraction removed [\%] & \num{62.6 \pm 0.7} & \num{86.4 \pm 0.7} & \num{99.2 \pm 0.4} & \num{99.2 \pm 0.4}& 100 & 100\\
       \hline
       Fit parameters & & & & & &\\
       Diffusion coeff. [$10^{-8}$ cm$^2$/sec] & $6.2^{+1.0}_{-4.0}$ & \num{10.7(0.4)} & \num{46.3(8.5)} & \num{45.3(6.5)} &\num{45(15)} & \num{39.4(7.6)}  \\ 
       Detrapping rate [$10^{-6}$/sec] & $2.0^{+2.6}_{-0.1}$ & $6.7^{+1.5}_{-0.8}$ & \num{357(52)} & \num{411(61)} & \num{334(71)} &  \num{344(55)}\\
       (Free + De-trapped) fraction [\%] & $62^{+1}_{-9}$ & \num{86 \pm 2} & \num{99 \pm 3} & $98^{+2}_{-17}$ & $99^{+2}_{-5}$ & $100^{+3}_{-7}$\\
       \hline
   \end{tabular}
   \caption{Summary of results for tritium removed from wafers at different temperatures. Upper Rows: Direct experimentally measured values. Residuals and fractions are calculated assuming all tritium is removed by the final bake at \SI{1000}{\celsius}. The upper limits on the residual activity are quoted at the 95\% CL. Lower Rows: Fit results for the diffusion and de-trapping parameters, including systematic uncertainties. See text for details on the model.}
   \label{tab:removal_frac}
\end{table*}
       
\subsection{Diffusion coefficient}
\label{sec:fitting}
To evaluate the diffusion coefficient we initially assumed, similarly to References \cite{ichimiya1968solubility, saeki1983origins}, that the tritium produced within the wafers is present in the silicon lattice as a mobile interstitial \cite{jones2008diffusion} and that release of tritium from the surface is relatively quick such that the time profile of the outgassed tritium is dominated by the diffusion of tritium through the bulk silicon. As will be discussed later, a more sophisticated model was required to match the experimental data, but we present this simplified model to better compare to previous works in literature.

\subsubsection{Simple diffusion model}
To model the diffusion analytically, we have assumed a semi-infinite slab of silicon of thickness $2a$, with the tritium concentration $c(x,t)$ [Bq/cm$^3$] initially uniform within the bulk and no tritium concentration outside the silicon. The Fick diffusion equation and boundary conditions are therefore
\begin{align}
&\frac{\partial c(x,t)}{\partial t}  = D \nabla^2 c(x,t)\\
&c(x, t=0) = c_0 & ~\text{for } -a < x < a\\
&c(a, t) = c(-a, t) = 0 & ~\text{for } t > 0
\end{align}
where $D$ [cm$^2$/sec] is the temperature-dependent diffusion coefficient of tritium in silicon. Solving these equations assuming a constant, spatially uniform temperature across the silicon, one can express the concentration in terms of a trignometric-series expansion\footnote{An equivalent solution can be expressed in terms of error functions $c(x,t) = c_0\left[ 1- \sum_{n=0}^\infty (-1)^n \erfc \left( \frac{(2n+1)a - x}{2\sqrt{Dt}}\right) - \sum_{n=0}^\infty (-1)^n \erfc \left( \frac{(2n+1)a + x}{2\sqrt{Dt}}\right) \right] $ which converges faster at short times \cite{crank1979mathematics}. However, the difference in the expressions for $n \ge 15$ terms was found to be negligible over the entire time span of our measurements.}
\begin{align}
c(x,t) &= \nonumber\\
& c_0\frac{4}{\pi}\sum_{n=0}^\infty\frac{(-1)^n}{2n+1}\cos\left(\frac{(2n+1)\pi}{2a}x\right)e^{-\left(\frac{(2n+1)\pi}{2a}\right)^2 Dt}
\label{eq:conc_simple}
\end{align}
The cumulative flux of tritium emitted from both surfaces, as a function of time, $M(t)$ [Bq/cm$^2$], can then be calculated from Fick's first law to be
\begin{align}
    M(t) &= c_02a\left[ 1- \frac{8}{\pi^2}\sum_{n=0}^\infty \frac{1}{(2n+1)^2} e^{-\left(\frac{(2n+1)\pi}{2a}\right)^2 Dt}\right] 
    \label{eq:cuml_trig}
\end{align}

\subsubsection{Data fitting with simple diffusion model}
\begin{figure*}[t]
    \centering
    \includegraphics[width=6in]{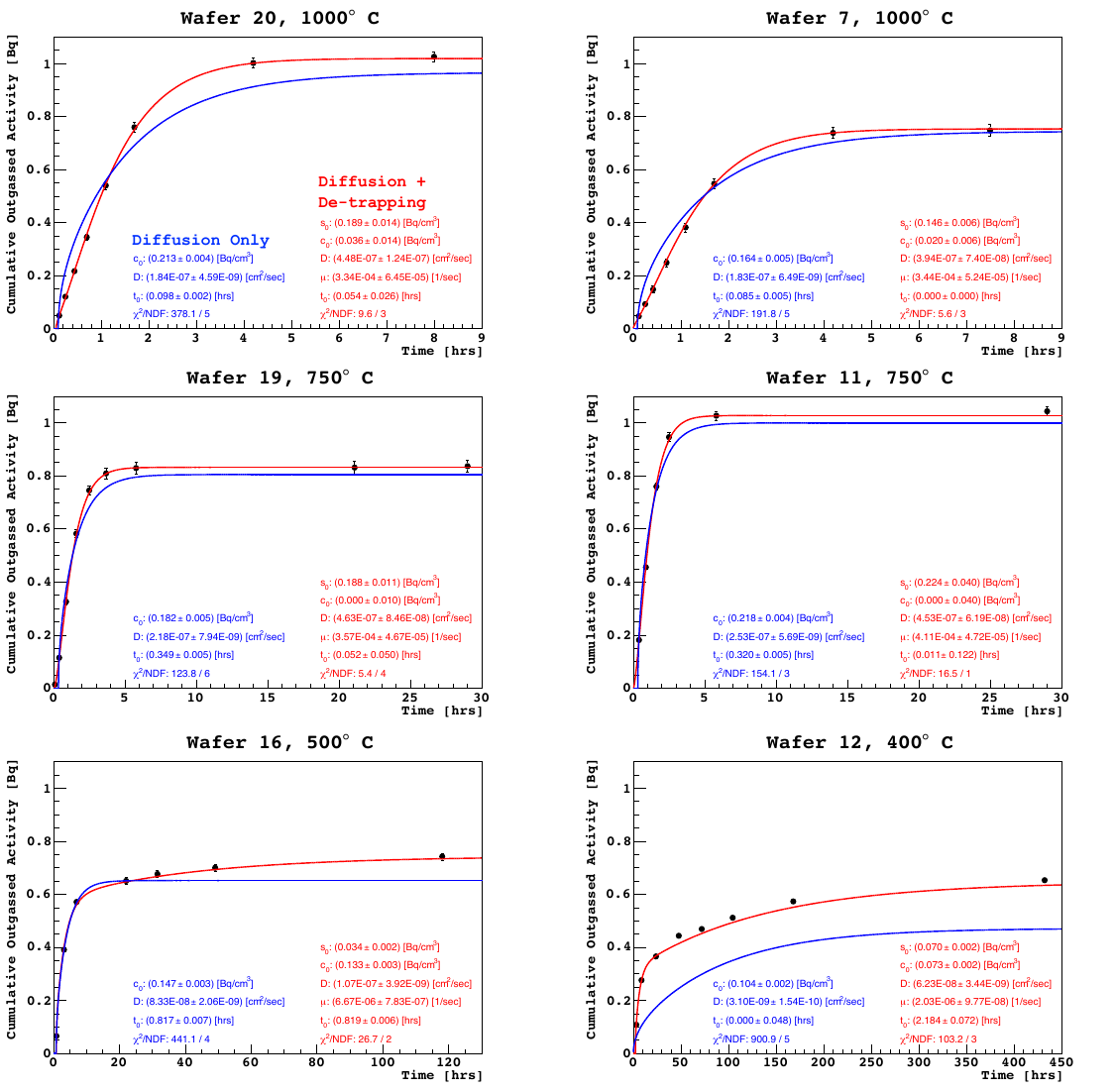}
    \caption{Comparison of measured cumulative outgassed activity to the best fit models. Fits were performed on the independent bubbler measurements (see Figure~\ref{fig:3H evo profiles}) but are shown in terms of the cumulative activity for easier visualization of trends. Blue: Fit to simple diffusion model. Red: Fit to model with diffusion and de-trapping. Note that the horizontal axes of the plots span different durations.}
    \label{fig:cuml_fits}
\end{figure*}
The data at each temperature (excluding the final measurement at $T=$ \SI{1000}{\celsius}) were fit to the simple diffusion model with the initial concentration $c_0$, diffusion constant $D$, and the initial start time $t_0$ left as free parameters, and the summation evaluated up to $n=100$. To avoid correlations between data points when fitting the cumulative outgassed tritium as a function of time, the individual bubbler measurements were fit to $M(t^i_{2}) - M(t^i_{1})$, where $t^i_{1}$ and $t^i_{2}$ are the start and end time of the $i^{th}$ bubbler collection respectively. For easier visualization of the outgassing trends, the best-fit model and the data are compared as a function of the cumulative outgassed flux (the cumulative outgassed flux at each bubbler change was calculated as the sum of all measured tritium measurements up until that time and the uncertainty calculated as the sum, in quadrature, of each of the independent measurements).  As can be seen from the fit results shown in blue in Figures~\ref{fig:3H evo profiles} and~\ref{fig:cuml_fits}, the simple diffusion curves do not match the data very well. We investigated whether the discrepancy at early times might be caused by a time lag in the wafers reaching the setpoint temperature but the measured time lag is not sufficient to account for the data time profile (see Appendix~\ref{sec:time_lag}). 

\subsubsection{Diffusion model with de-trapping}
\label{sec:detrap_model}
As mentioned earlier, hydrogen isotopes can form complexes with impurities and defects that are stable at lower temperatures and can lead to trapping and lower ``effective diffusivity''~\cite{sopori2001silicon, pearton1987hydrogen}. To better match the data, we developed a model to include the effects of some tritium trapped at defect sites within the silicon lattice. The trapped tritium is considered immobile, but tritium that is less tightly bound to defects can de-trap at elevated temperatures and enter the interstitial space. We assume a first-order de-trapping reaction with a rate proportional to the de-trappable concentration. The equations that govern the free interstitial tritium concentration ($c(x,t)$ [Bq/cm$^3$]) and the concentration of trapped tritium that can de-trap ($s(x,t)$ [Bq/cm$^3$]) are given by
\begin{align}
\label{eq:fick1}
&\frac{\partial c(x,t)}{\partial t}  = D \nabla^2 c(x,t) - \frac{\partial s(x,t)}{\partial t}\\
\label{eq:fick2}
&\frac{\partial s(x,t)}{\partial t}  = - \mu s(x,t)\\
&c(x, t=0) = c_0 & ~\text{for } -a < x < a\\
&c(a, t) = c(-a, t) = 0 & ~\text{for } t > 0\\
&s(x, t=0) = s_0 & ~\text{for } -a < x < a
\end{align}
where $\mu$ [1/sec] is the rate of detrapping. The solution of these equations (as well as a more general model that includes trapping) is outlined in Appendix~\ref{sec:laplace}, and here we simply present the results
\begin{align}
    \label{eq:desorbsolnfree}
c(x,t) =&~\frac{4c_0}{\pi}\sum_{n=0}^{\infty}\frac{(-1)^n\cos{\left(\frac{(2n+1)\pi}{2a}x\right)}e^{{p_n}t}}{(2n+1)} \nonumber \\ 
&+ \frac{4s_0}{\pi}\sum_{n=0}^{\infty}\frac{(-1)^n\cos{\left(\frac{(2n+1)\pi}{2a}x\right)}e^{{p_n}t}}{(2n+1)(p_n/\mu+1)} \nonumber \\
&+ s_0e^{-\mu t}\left(\frac{\cos{(\sqrt{\frac{\mu}{D}} x)}}{\cos{(\sqrt{\frac{\mu}{D}} a)}}  - 1 \right)\\
\label{eq:desorbsolnfixed}
s(x,t) =&~s_0e^{-\mu t}\\
\label{eq:desorbsolnpn}
\text{where } p_n &= \frac{-(2n+1)^2\pi^2D}{4a^2}
\end{align}
The cumulative tritium flux removed from the silicon is given by
\begin{align}
M(t) =& ~(c_0+s_0)2a \nonumber \\ 
&- \frac{16a}{\pi^2}\sum_{n=0}^{\infty}\frac{(c_0 + s_0\mu/(p_n+\mu))e^{{p_n}t}}{(2n+1)^2} \nonumber\\
&- 2s_0\sqrt{\frac{D}{\mu}}\tan{\left(\sqrt{\frac{\mu}{D}} a\right)}e^{-\mu t}
\label{eq:cuml_outgas}
\end{align}
\subsubsection{Data fitting with diffusion and detrapping model}
The data were fit to the model with de-trapping, with the initial free ($c_0$) and de-trappable ($s_0$) concentrations, the diffusion constant ($D$), the de-trapping rate ($\mu$), and the initial start time ($t_0$) left as free parameters. The best-fit results are shown in red in Figures~\ref{fig:3H evo profiles} and~\ref{fig:cuml_fits}. It can be seen that the model including de-trapping matches the data much better than the simple diffusion model. As discussed in Appendix~\ref{sec:laplace}, the time profile of the removed tritium depends on the relative magnitude of the diffusion ($D/a^2$) and de-trapping rates. At high temperatures ($\geq$ \SI{750}{\celsius}) the de-trapping rate is fast compared to the diffusion, and after a relatively short initial period all the tritium is free and is removed following a simple diffusion curve. At lower temperatures the de-trapping rate is slow compared to the diffusion, leading to a slower emission of trapped tritium even at later times. At the lowest temperature (\SI{400}{\celsius}) the de-trapping model still does not match the data very well, perhaps indicating additional factors not included in the model (see Section~\ref{sec:discussion} for further discussion). We note that the fit values for the initial free and de-trappable concentrations are highly negatively correlated at the higher temperatures (correlation coefficient $< -0.9$) with the best-fit values not well-constrained. However, the sum of the two fit components (including correlations) as a fraction of the measured total activity of each wafer, $A_{tot}$, shows good agreement with the directly measured values (see Figure~\ref{fig:frac_removed}) indicating that the model predicts not much additional tritium can be extracted at each temperature, beyond the experimentally measured values. Systematic uncertainties were evaluated by looking at the range of variation when varying the start and end points of the fits and including a re-trapping term. The fit results, combining statistical and systematic uncertainties, for the diffusion constant, de-trapping rate, and the combined concentration fraction are shown in Table~\ref{tab:removal_frac}.

\subsection{Temperature Dependence}
\subsubsection{Diffusion vs Temperature}
\begin{figure}[t]
    \centering
    \includegraphics[width=1\linewidth]{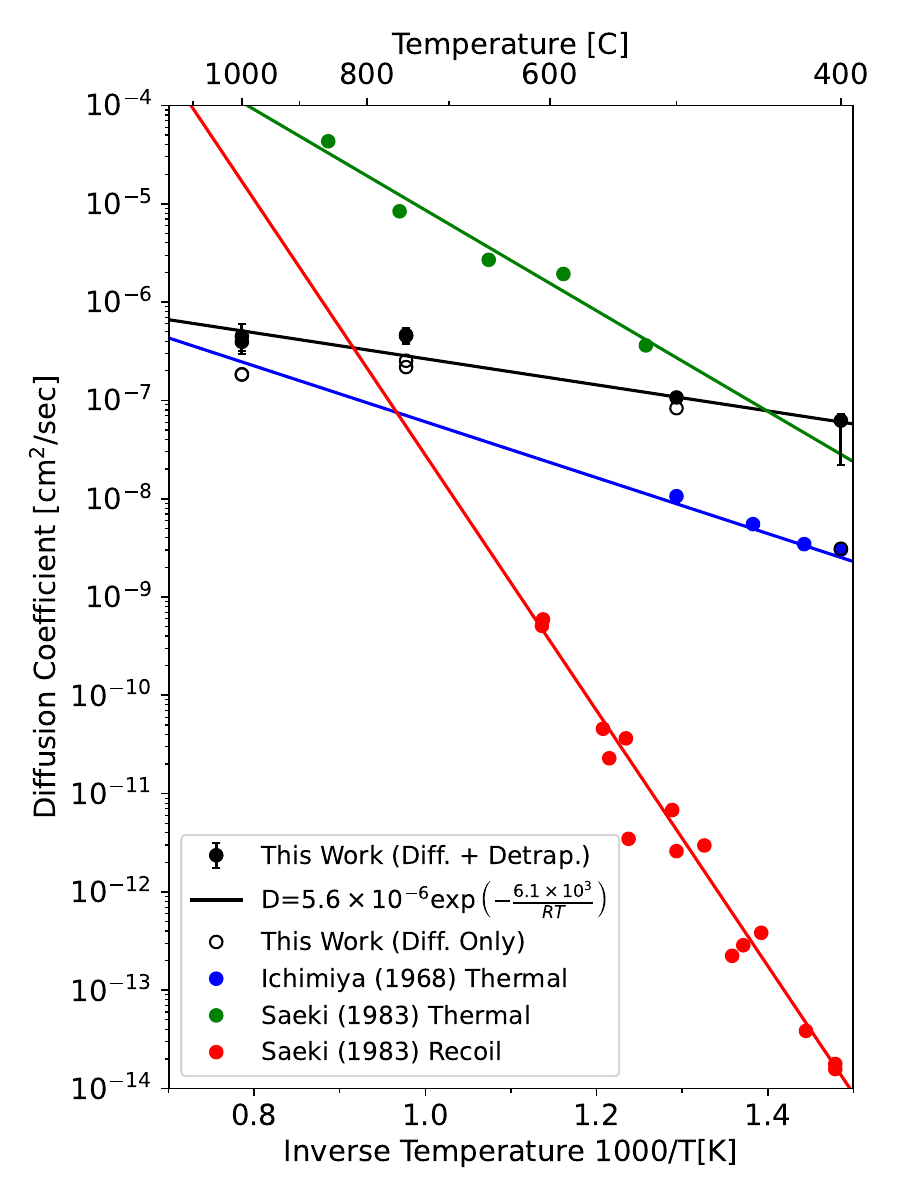}
    \caption{Fitted diffusion coefficient versus inverse temperature, compared to previous results in the literature \cite{ichimiya1968solubility, saeki1983origins}. The filled circles show the results from the fit with the diffusion and de-trapping model, while the empty circle shows the fit to a simple diffusion model (central value only). The black like is a fit to the filled data points with the Arrhenius equation.}
    \label{fig:diff_results}
\end{figure}

The temperature dependence of the diffusion coefficient in solids generally follows the Arrhenius equation
\begin{align}
    D(T) = D_0 \exp{\left(-\frac{E_D}{RT}\right)}
    \label{eq:d_vs_T}
\end{align}
where $E_D$ [cal/mol] is the activation energy for diffusion and $R \sim$ \SI{1.987}{cal \per \kelvin \per mol} is the molar gas constant. Figure~\ref{fig:diff_results} shows the fitted diffusion coefficient vs temperature along with other previous measurements of tritium diffusion in silicon. For comparison to the other measurements in the literature that used a simple diffusion model, we also show the central values of the simple diffusion fit, even though it is not a good match to our data. We have fit our diffusion and de-trapping model results with Eqn.~\ref{eq:d_vs_T} (black line in figure), which gives values of $D_0 =$ \SI{5.6(1.7)E-6}{\cm\squared\per\second} and $E_D = $ \SI{6.1(0.5)E3}{cal/mol}.

\subsubsection{De-trapping vs Temperature}
\begin{figure}[t]
    \centering
    \includegraphics[width=1\linewidth]{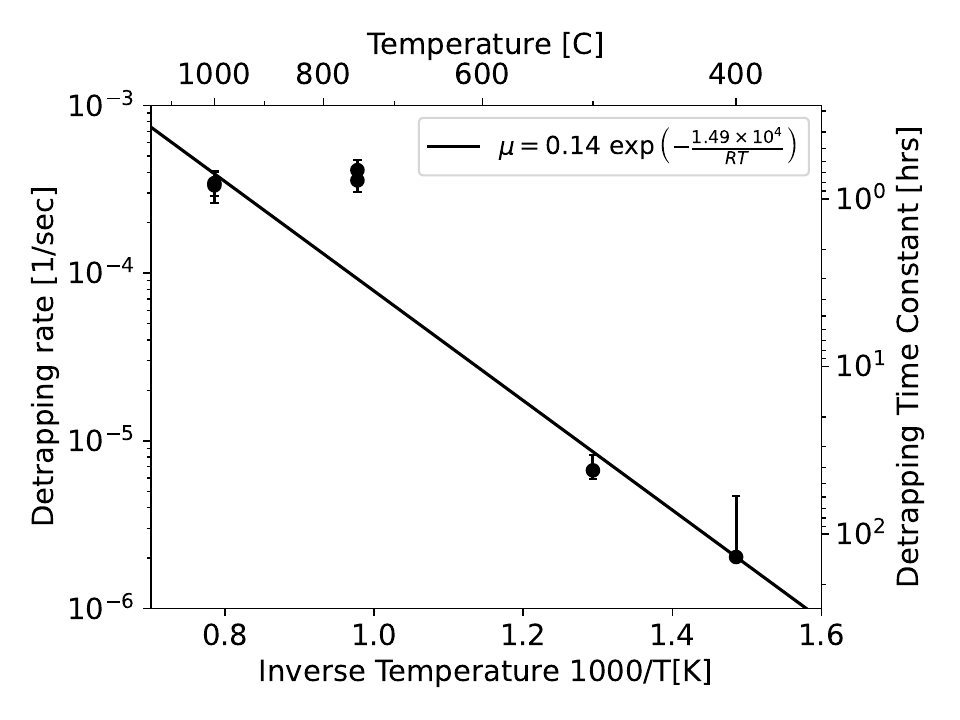}
    \caption{Fitted de-trapping rate versus inverse temperature. The black like is a fit to the filled data points with the Arrhenius equation. See text for fit parameters.}
    \label{fig:detrap_rate}
\end{figure}

If the $\mu$ parameter in our model does in fact represent a first-order rate constant for tritium being released from defects, one would expect it to also follow the Arrhenius equation as a function of temperature. In Figure~\ref{fig:detrap_rate} we show the fitted de-trapping parameter $\mu$ versus temperature. Fitting the temperature dependence with the functional form of Eq.~\ref{eq:d_vs_T}, we obtain values of $\mu_0$ = \SI{0.14 \pm 0.03}{\per\second} and $E_\mu$ = \SI{1.49(0.03)E4}{cal/mol}. Although the fit to the Arrhenius equation is not very good, this de-trapping activation energy, \SI{0.65 \pm 0.02}{\eV}, is similar to previous measurements of the de-trapping rate of deuterium in silicon, which found the dominant de-trapping activation energy to be \SI{1.2 \pm 0.1}{\eV}, along with smaller contributions  at 0.33 and 1.8 eV \cite{abrefah1990hydrogen}. It is also on the same order of magnitude as the calculated bonding energies of atomic hydrogen (protium) at different sites within a silicon lattice (for example, see Ref.~\cite{tuttle1999structure} which estimates the Si-H dissociation activation energy to be less than 3 eV).

\subsubsection{High temperature behavior}
It can be seen that both the diffusion constant and de-trapping rate at \SI{750}{\celsius} are consistent with the values measured at \SI{1000}{\celsius}. This is not an artifact of the fit; the measured time profiles at those two temperatures are very similar, across all four measurements at these temperatures - see Section~\ref{sec:discussion} for further discussion.

\section{Discussion}
\label{sec:discussion}
Relevant to low-radioactivity applications that aim to reduce cosmogenic tritium backgrounds in silicon substrates, our results show that tritium produced by cosmic-ray-like neutron spallation interactions in high-purity silicon can be removed from the silicon substrate at elevated temperatures. We have demonstrated complete removal of tritium at $\geq$ \SI{750}{\celsius}. This is similar to the behavior observed with natural hydrogen introduced into silicon, which is attributed to the breaking of Si-H bonds at temperatures around \SIrange{700}{800}{\celsius}  \cite{stein1975bonding, brodsky1977quantitative, pearton1987hydrogen, keinonen1988defect}. The fraction of tritium that is removable from the silicon decreases with decreasing temperature, which indicates that some of the tritium is bound within the silicon lattice and is not mobile at low temperatures. However a significant reduction in tritium backgrounds can be achieved even at moderate baking temperatures (roughly 63\% at \SI{400}{\celsius}).

 In addition to the removal fractions, the experimental method presented in this paper allowed for measurement of the time evolution of the tritium outgassing, which can be used to model the behavior of tritium and hence extrapolate removal times to different substrate thicknesses and improve our understanding of hydrogen transport in crystalline silicon. The time profiles of tritium outgassing from our irradiated silicon samples are not well described by a simple diffusion model. However, by introducing a first-order de-trapping parameter in addition to free diffusion we are able to accurately model the experimental data and extract the temperature-dependent diffusion and de-trapping parameters. 
 
 Our fitted diffusion coefficients lie roughly between the previous measurements of thermally-introduced tritium in silicon obtained by Ichimiya and Furuichi \cite{ichimiya1968solubility} and Saeki et. al. \cite{saeki1983origins}. Those previous measurements heated the silicon samples in molecular hydrogen (tritium and protium), reaching concentrations of \SIrange{1E11}{1E14}{atoms\per\cm\cubed} \cite{ichimiya1968solubility, saeki1983origins}. In our samples, nuclear spallation generates atomic tritium, and the much lower concentrations ($\sim$\SI{3E8}{atoms\per\cm\cubed}) makes it more unlikely to combine to form molecular tritium, which is significantly less mobile than atomic tritium at lower temperatures \cite{pearton1987hydrogen} and may contribute to the weaker temperature dependence of our measured diffusion coefficient. For comparison, a silicon sample exposed to natural cosmic rays for a year would only have an atomic tritium concentration of $\sim$\SI{100}{atoms\per\cm\cubed}. 
 
 Our results are orders of magnitude larger than the recoil-implanted tritium measurements by Saeki et. al. \cite{saeki1983origins}. The reason for this discrepancy is not understood. One of the hypothesized reasons for why the recoil-driven samples showed significantly retarded diffusion compared to the thermally-driven samples in the Saeki  et. al. measurements was the production of localized defects in the crystal lattice during recoil implantation \cite{saeki1983origins}. However, spallation-induced tritium production also produces significant radiation damage in the lattice. In addition to the damage caused by the incident high-energy neutrons, Geant4 simulations of the triton particles produced by the LANSCE neutron beam estimate a median triton kinetic energy of roughly \SI{11}{\MeV}, significantly higher than the  \SI{2.7}{\MeV} and \SI{191}{\keV} recoils from the $^6$Li(n,$\alpha$)T and $^3$He(n,p)T reactions used in Ref.~\cite{saeki1983origins}. We note that for the recoil-injected tritium measurements, Saeki et. al. used an approximate linear relationship between the fractional release rate of tritium and the square root of the annealing time to determine the diffusion coefficients, assuming that diffusion is the rate-determining step in the release of tritium. However, even when our data are fit to a simple diffusion model (using the full analytical solution) we do not obtain diffusion coefficients that are as small as those measured by Saeki et. al. It should be pointed out that the wafers in this measurement were exposed to a neutron fluence of roughly \SI{8E12}{neutrons} above \SI{10}{\MeV}, which corresponds to roughly \num{7E7} years of sea-level exposure. Silicon substrates used for rare event searches will typically have less than a year of sea-level exposure and hence significantly less cosmic-ray induced radiation damage, making it less likely for the diffusion to be impeded by trapping of interstitial tritium on defects.
 
 As noted earlier, our single rate de-trapping model does not match the data very well at the lowest (\SI{400}{\celsius}) temperature, leading to large uncertainties in the estimated parameters. This is likely because at these lower temperatures there are several different trapping sites that are active, each with their own characteristic energy that can trap and release the tritium as it migrates through the lattice. Evidence for this can be seen for example by the Si-H IR bands in neutron-irradiated crystalline silicon shown in Figure 46 of Ref.~\cite{pearton1987hydrogen}. At higher temperatures many of these damage-produced IR bands anneal out, leading to behavior that is simpler to model. 
 
 The close similarity of the tritium outgassing profiles at $750 \degree$ C and $1000 \degree$ C indicates a possible departure from the typical Arrhenius rate dependence at high temperatures. While such behavior has been observed in other materials, this is in tension with previous measurements and models of hydrogen (protium) diffusion in silicon \cite{bedard2000diffusion} and needs additional temperature measurements to resolve.  

In summary, the measured removal fractions, diffusion, and de-trapping coefficients in this work imply that any tritium generated by cosmic ray exposure of high-purity silicon can be completely removed by baking the silicon at $\geq 750\degree$ C for a relatively short time period. For example, for the 200-mm wafers planned for the next-generation Oscura CCD dark matter experiment, $> 95\%$ of the cosmogenic tritium can be removed by baking the \SI{725}{\micro\meter}-thick wafers for 2.5 hours at $1000\degree$ C. This greatly reduces the constraints on above-ground exposure of silicon intended for use in low-background rare event searches before device fabrication, including the growing of the silicon boule, transportation, wafering, etc. 

\section{Future work}
Given that cosmogenic tritium can be removed from pre-fabricated pure silicon, future work will focus on mitigating the tritium background from cosmogenic exposure during device fabrication, which often happens in un-shielded above-ground facilities. The fabrication of devices and detectors from silicon wafers typically involves several steps, including the addition of thin conductive, insulating, ion-implanted, and passivation layers, which are then patterned by microlithographic processes to obtain the final device structure (see for example the CCD \cite{janesick2001scientific, Holland:2003kiw} or SuperCDMS detector \cite{jastram2015cryogenic} fabrication process). The details of the process vary depending on the specific device and some fabricated devices may not be able to withstand extended periods of time at elevated temperatures, making removal of the tritium generated after device fabrication unfeasible. We note that the CCD fabrication process already involves several high temperature ($\geq 750\degree$ C) intermediate steps \cite{janesick2001scientific} that could be extended to remove tritium, though it also involves adding several additional layers (nitride, oxide, metallization) that could impede the diffusion out of the silicon. We are currently investigating whether tritium is able to efficiently diffuse through the various CCD surface layers and hence if tritium generated by cosmic ray exposure during fabrication may also be at least partially removed. 

\section{Acknowledgments}
We would like to thank Gabe Ortega at PNNL for designing the wafer holders for the beam exposure. We are grateful to Liang Yang, William Thompson, Stephen Wender, Kranti Gunthoti, and Larry Rodriguez for technical assistance during the LANSCE beam exposures. We also thank Cory Overman at PNNL for guidance on laser dicing of the silicon wafers. We would like to acknowledge Jonathan Burnett, Allan Myers, and Manish Sharma for measuring the wafers on the SAGE Ge detector at PNNL.

This work was supported by the DOE Office of Science as part of the Office of High Energy Physics Cosmic Frontier research program at Pacific Northwest National Laboratory (PNNL), a multi-program national laboratory operated for the U.S. Department of Energy (DOE) by Battelle Memorial Institute under Contract No. DE-AC05-76RL01830. The work at University of Chicago and University of Washington was supported through Grant No. NSF PHY-2413013 and 2413014. This work was supported by the Kavli Institute for Cosmological Physics at the University of Chicago through an endowment from the Kavli Foundation.
 
\appendix
\section{Discrepancy at early times}
\label{sec:time_lag}

\begin{figure}[t]
    \centering
    \includegraphics[width=1\linewidth]{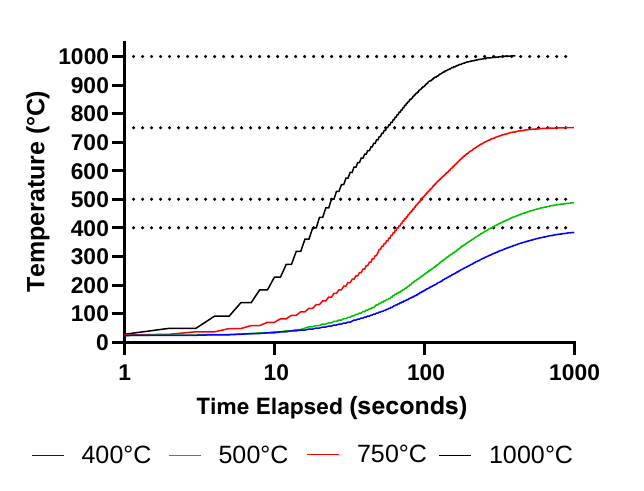}
    \caption{Sample heating lag time. The accuracy of the thermocouple measurements is ± 0.25\%}
    \label{fig:lag time}
\end{figure}

\begin{table}[t!]
   \centering
   \begin{tabular}{l|cccc} % Column formatting, @{} suppresses leading/trailing space
       & \SI{400}{\celsius} & \SI{500}{\celsius} & \SI{750}{\celsius} & \SI{1000}{\celsius} \\
        \hline
       \multirow{2}{1.15 in}{Time to reach 95\% of setpoint [sec]} &  \multirow{2}{*}{844} & \multirow{2}{*}{667} & \multirow{2}{*}{285} & \multirow{2}{*}{138}\\
       & & & &\\
       \multirow{2}{1.15 in}{First bubbler measurement [sec]} & \multirow{2}{*}{11220} & \multirow{2}{*}{3180} & \multirow{2}{*}{1440} & \multirow{2}{*}{480} \\
        & & & &\\
   \end{tabular}
   \caption{Comparison of time taken for quartz boat to reach the setpoint temperature with the time of the first bubbler measurement.}
   \label{tab:lag_times}
\end{table}

One possible explanation for the difference in the data and the diffusion model at early times is that the model assumes a constant temperature for the wafer, while in reality the wafers are initially at room temperature and then heated to the set temperature in the Pyrolyser. The lower temperatures at early times could explain the slower than expected outgassing of tritium. To explore this possibility,  the time required for the quartz boat and the sand (used to support the wafer) to equilibrate with the set temperature was measured. A K-type thermocouple was placed so that the junction end was partially in the sand bed and partially exposed to the gas flow (to replicate the wafer position in the quartz boat). While no silicon wafers were included in the test, the time required for the wafer bulk to reach equilibrium with the surface, given the wafer thickness (\SI{1}{mm}) and the thermal diffusion constant of silicon (\SI{88}{\mm\squared\per\second}), should be negligible. The thermocouple was attached to a data logger and subsequently downloaded to a PC for processing (see Figure \ref{fig:lag time}). 

As can be seen from the values in Table~\ref{tab:lag_times}, the time required to thermally equilibrate is significantly shorter than the collection time of the first bubbler, and so we do not expect any effect on later bubbler measurements. 

\begin{figure}
    \centering
    \includegraphics[width=\linewidth]{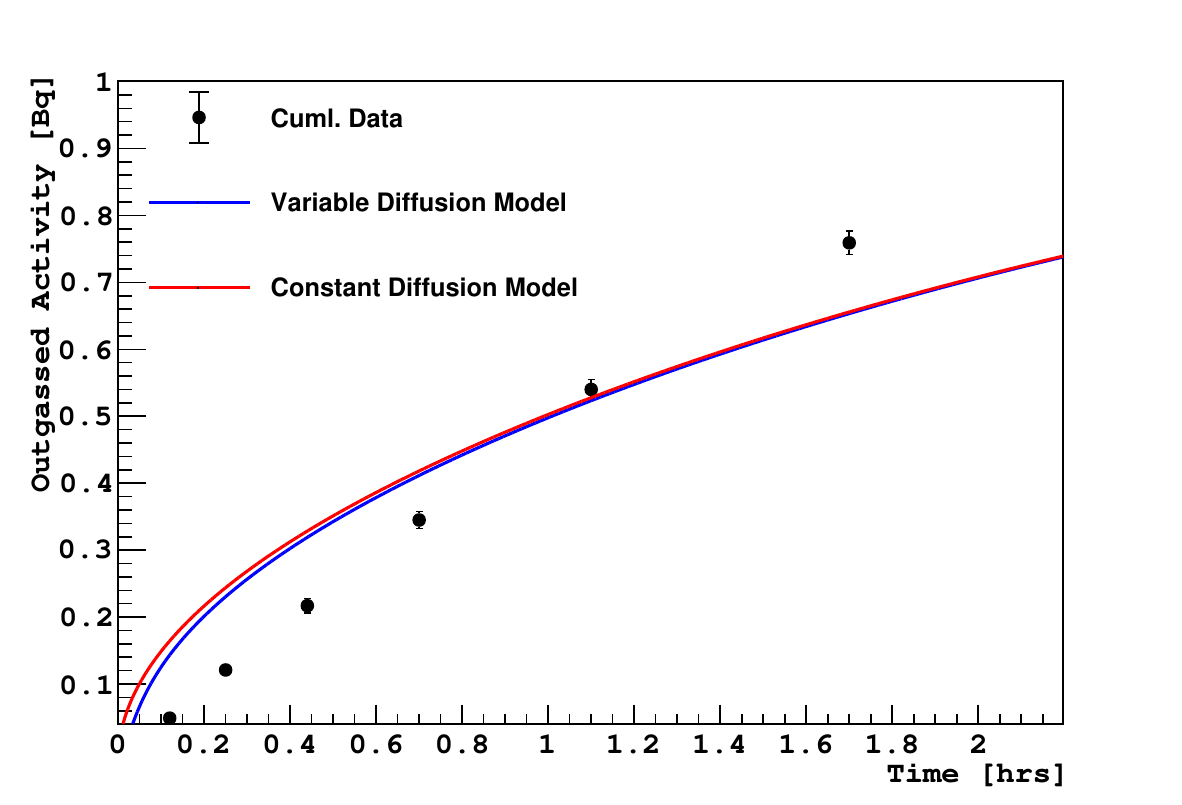}
    \caption{Fit results to outgassing data from wafer 20 at \SI{1000}{\celsius}, keeping the starting time fixed to zero. The continuous lines show the fit with a fixed (red) and temperature-dependent (blue) diffusion constant.}
    \label{fig:time_var_diff}
\end{figure}

To confirm that the thermal equilibration time has a small effect on the measurement, one can solve Fick's equations for a time-varying diffusion coefficient $D(t)$ by making a transformation of the time variable
\begin{align}
    s(t) \equiv \int_0^t{\frac{D(t)}{D_\infty}} dt
\end{align}
where $D_\infty$ is the diffusion coefficient at the equilibrium temperature \cite{crank1979mathematics}. With this transformation, we recover the original Fick equation in one dimension
\begin{align}
\frac{\partial c(x,s)}{\partial s} = D_\infty \frac{\partial^2 c(x,s)}{\partial x^2}
\end{align}
which has the same solutions Eqs~\ref{eq:conc_simple} and~\ref{eq:cuml_trig}. To make the transformation, one needs to define how the diffusion coefficient varies with temperature. For this cross-check, we assumed that the diffusion coefficient follows the standard Arrhenius behavior (Eq.~\ref{eq:d_vs_T}) where the time dependence of the temperature ($T(t)$) was taken from the thermocouple measurements shown in Figure~\ref{fig:lag time} and the activation energy was taken from Ichimiya and Furuichi \cite{ichimiya1968solubility} who measured $E_a =$ \SI{13000}{cal/mol}. The fit to the \SI{1000}{\celsius} data (where the time lag is most likely to have an effect) with both a fixed and time-varying diffusion coefficient (keeping the start time fixed) is shown in Figure~\ref{fig:time_var_diff}. As can be seen, the difference between the models is minimal and is not sufficient to match the data at early times. As a final cross-check, we used the COMSOL Multiphysics \cite{comsol} software package to model the time-varying diffusion in the silicon wafers, using the measured temperature time profiles in  Figure~\ref{fig:lag time}. At all temperatures the differences between the fixed and time-varying temperature model were small and could not account for the trends in the data. 

\section{Analytical diffusion model with trapping and de-trapping}
\label{sec:laplace}
We consider the scenario in which tritium can be trapped on stationary defects in the silicon lattice. These defects can be pre-existing or created during the spallation event and subsequent stopping of the triton nucleus in the silicon lattice. We will assume that at any given fixed temperature there is a rate of absorption of free, interstitial tritium at defect sites that is proportional to the concentration of free tritium and a rate of desorption of trapped tritium into the interstitial space (where it is free to diffuse) that is proportional to the concentration of trapped tritium. We note that in reality there could be several different types of defects, each with their own characteristic temperature-dependent trapping and de-trapping rate constants, but for simplicity we only consider a single defect type here.

 Modifying Fick's second law to account for absorption and desorption, we have
\begin{align}
\frac{\partial c(x,t)}{\partial t}  &= D \nabla^2 c(x,t) - \frac{\partial s(x,t)}{\partial t}\\
\frac{\partial s(x,t)}{\partial t}  &= \lambda c(x,t) - \mu s(x,t)
\end{align}
where $c(x,t)$ is the time and spatially dependent concentration [atoms/cm$^3$] of free, interstitial tritium, $s(x,t)$ is the time and spatially dependent concentration [atoms/cm$^3$] of trapped tritium, $D$ is the diffusion coefficient [cm$^2$/sec], $\lambda$ [1/sec] is the rate of absorption, and $\mu$ [1/sec]  is the rate of desorption.
For tritium initially distributed uniformly throughout a semi-infinite slab of silicon with thickness $2a$, and the concentration on the outside of the slab held at zero (infinite flushed volume), we have the following boundary conditions
\begin{subequations}
\label{eq:bc}
\begin{align}
\label{eq:bc1}
c(x, t=0) &= c_0 & ~\text{for } -a < x < a\\
\label{eq:bc2}
c(a, t) &= 0 & ~\text{for } t > 0 \\
\label{eq:bc3}
c(-a, t) = 0 &\text{ or } \left.\frac{\partial c(x,t)}{\partial x}\right|_{x=0} = 0 & ~\text{for } t > 0 \\
\label{eq:bc4}
s(x, t=0) &= s_0 & ~\text{for } -a < x < a
\end{align}
\end{subequations} 
We will follow the methods outlined in Ref.~\cite{crank1979mathematics} Sec. 14.4 and solve this by taking the Laplace transform of the differential equations, including boundary conditions ~\ref{eq:bc1} and ~\ref{eq:bc4}
\begin{align}
\label{eq:lp1}
&D\frac{\partial^2 \tilde{c}(x,p)}{\partial x^2} - p(\tilde{c}(x, p) + \tilde{s}(x, p))+ c_0+ s_0 = 0\\
\label{eq:lp2}
&\tilde{s}(x, p)(p+ \mu) = \lambda \tilde{c}(x,p) + s_0
\end{align}
Similarly transforming the remaining unused boundary conditions
\begin{subequations}
\begin{align}
\label{eq:lpbc1}
\tilde{c}(a,p) &= 0 \\
\label{eq:lpbc2}
\left.\frac{\partial \tilde{c}(x,p)}{\partial x}\right|_{x=0} &=  0 
\end{align}
\end{subequations}
One can now solve the second order linear ordinary differential equations to get
\begin{align}
\tilde{c} &= \frac{j(p)}{k^2(p)} \left(1-\frac{\cos{(kx)}}{\cos{(ka)}}\right)\\
k^2(p) &\equiv - \frac{p(\lambda + p + \mu)}{D(p+\mu)}\\
j(p) &\equiv -\frac{(c_0(p+\mu) + s_0\mu)}{D(p+\mu)}\\
\tilde{s} &= \frac{\lambda \tilde{c} + s_0}{p+\mu}
\end{align}
To take the inverse Laplace transform of these solutions we will use the convenient result quoted in Ref.~\cite{crank1979mathematics} Eq. 2.59 for the inverse Laplace transform of a ratio of two polynomials (under certain conditions)
\begin{align}
\mathcal{L}{^{-1}}\left\{\frac{f(p)}{g(p)}\right\} &= \sum_{r}\frac{f(p_r)e^{{p_r}t}}{g'(p_r)} 
\end{align}
where $p_r$ are the roots of $g(p)$ and $g'(p_r) \equiv \left .\frac{dg}{dp}\right|_{p=p_r}$. Using this result we find the solutions for the free and fixed concentrations of tritium as a function of position and time
\begin{align}
\label{eq:finalfree}
&c(x,t) = \nonumber \\&\frac{4}{\pi}\sum_{n=0}^{\infty}\frac{(-1)^n(p_n+\mu)(c_0(p_n+\mu) + s_0\mu)\cos{\left(\frac{(2n+1)\pi}{2a}x\right)}e^{{p_n}t}}{(2n+1)(\lambda\mu + (p_n+\mu)^2)}\\
\label{eq:finalfixed}
&s(x,t) = \nonumber \\&\frac{4\lambda}{\pi}\sum_{n=0}^{\infty}\frac{(-1)^n(c_0(p_n+\mu) + s_0\mu)\cos{\left(\frac{(2n+1)\pi}{2a}x\right)}\left(e^{p_n t}-e^{-\mu t}\right)}{(2n+1)(\lambda\mu + (p_n+\mu)^2)} \nonumber\\&  + s_0e^{-\mu t}\\
\label{eq:finalpn}
&\text{where } - \frac{p_n(\lambda + p_n + \mu)}{D(p_n+\mu)} = \frac{(2n+1)^2\pi^2}{4a^2}
\end{align}
Note that in general the solutions above each have two infinite series corresponding to the two roots of Eq.~\ref{eq:finalpn} for $p_n$. The cumulative outgassed flux of tritium at any time can be calculated by evaluating the difference between the initial amount of tritium and the current amount of tritium (both fixed and free)
\begin{align}
&M(t) \equiv \int_{-a}^{+a}{(c(x,0) + s(x,0))dx} - \int_{-a}^{+a}{(c(x,t) + s(x,t))dx} \nonumber \\
&=  (c_0+s_0)2a - \frac{16a}{\pi^2}\sum_{n=0}^{\infty}\frac{(p_n+\mu+\lambda)(c_0(p_n+\mu) + s_0\mu)e^{{p_n}t}}{(2n+1)^2(\lambda\mu + (p_n+\mu)^2)} \nonumber\\
&- 2ae^{-\mu t} \left(s_0 - \frac{8\lambda}{\pi^2}\sum_{n=0}^{\infty}\frac{(c_0(p_n+\mu) + s_0\mu)}{(2n+1)^2(\lambda\mu + (p_n+\mu)^2)} \right)
\end{align}
The numerical evaluation of these equations is complicated by the slow convergence of these equations for certain situations (see discussion in Ref.~\cite{crank1979mathematics} Section 14.4) and including the trapping rate parameter $\lambda$ did not have any appreciable effect on the fit other than at the lowest (400\degree C) temperature. Here we focus on two simplifications to allow us to easily fit the experimental data.
\subsection{Simple diffusion model}
We can check what the equations reduce to if there is no absorption or desorption to or from trapped states, i.e. $\lambda = \mu = s_0 = 0$.
Plugging this first into the equation for $p_n$ (Eq.~\ref{eq:finalpn})
\begin{align}
p_n &= \frac{-(2n+1)^2\pi^2D}{4a^2} 
\end{align}
We can then calculate the concentration of the free term (there is no fixed concentration) from Eq~\ref{eq:finalfree}
\begin{align}
\label{eq:simplediffusion}
c(x,t) &= \frac{4c_0}{\pi}\sum_{n=0}^{\infty}\frac{(-1)^n \cos{\left(\frac{(2n+1)\pi}{2a}x\right)}e^{{\frac{-(2n+1)^2\pi^2D}{4a^2}}t}}{(2n+1)}
\end{align}
which is the well-known result for simple diffusion (see e.g. Ref.~\cite{crank1979mathematics} Eq. 2.67). Similarly, for the cumulative outgassed flux 
\begin{align}
M(t) &= c_02a\left(1 - \frac{8}{\pi^2}\sum_{n=0}^{\infty}\frac{e^{{\frac{-(2n+1)^2\pi^2D}{4a^2}}t}}{(2n+1)^2}\right)
\end{align}
\subsection{Diffusion and de-trapping model}
For the case where trapping of free interstitial tritium at defect sites is extremely unlikely, i.e. $\lambda \to 0$, due to a low concentration of pre-existing, unoccupied defects, we only have to account for the de-trapping of tritium from defect sites. In this case, defect sites that have a characteristic trapping energy well above the set temperature will retain the trapped tritium indefinitely and are effectively de-coupled, while $s_0$ now refers to only the ``de-trappable'' component where the characteristic energy is comparable to the thermal energy. We can similarly solve the differential equations using Laplace transformations to obtain
\begin{align}
c(x,t) =&~\frac{4c_0}{\pi}\sum_{n=0}^{\infty}\frac{(-1)^n\cos{\left(\frac{(2n+1)\pi}{2a}x\right)}e^{{p_n}t}}{(2n+1)} \nonumber \\ 
&+ \frac{4s_0}{\pi}\sum_{n=0}^{\infty}\frac{(-1)^n\cos{\left(\frac{(2n+1)\pi}{2a}x\right)}e^{{p_n}t}}{(2n+1)(p_n/\mu+1)} \nonumber \\
&+ s_0e^{-\mu t}\left(\frac{\cos{(\sqrt{\frac{\mu}{D}} x)}}{\cos{(\sqrt{\frac{\mu}{D}} a)}}  - 1 \right)\\
s(x,t) =&~s_0e^{-\mu t}\\
\text{where } p_n &= \frac{-(2n+1)^2\pi^2D}{4a^2}
\end{align}
In this case the cumulative tritium flux removed from the silicon is given by
\begin{align}
M(t) =& ~(c_0+s_0)2a \nonumber \\ 
&- \frac{16a}{\pi^2}\sum_{n=0}^{\infty}\frac{(c_0 + s_0\mu/(p_n+\mu))e^{{p_n}t}}{(2n+1)^2} \nonumber\\
&- 2s_0\sqrt{\frac{D}{\mu}}\tan{\left(\sqrt{\frac{\mu}{D}} a\right)}e^{-\mu t}
\end{align}

\begin{figure}[t]
   \centering
   \includegraphics[width=0.49\textwidth]{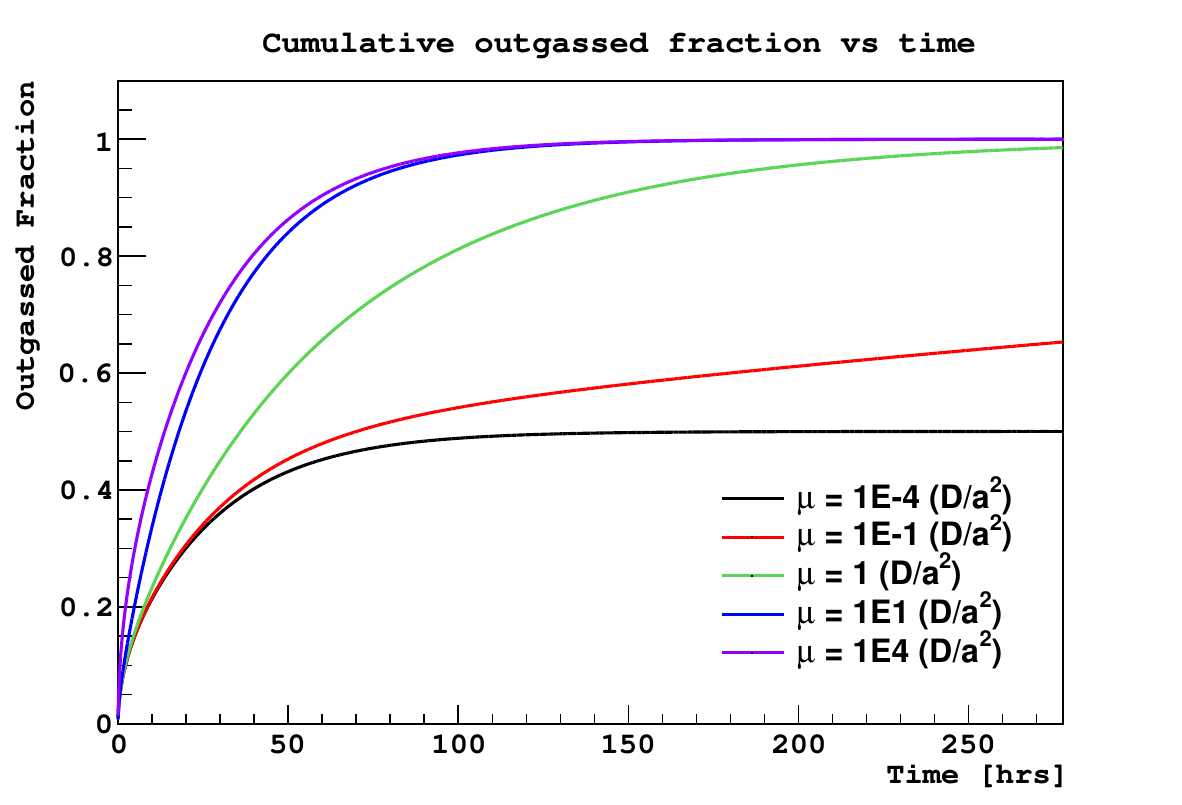}
   \caption{Numerically evaluated cumulative outgassed fraction, $M(t)/M(\infty)$, for different detrapping rates. We have used $a = 0.05$ cm, $D = 1\times 10^{-8}$ cm$^2$/sec, $c_0 = s_0 = 1$ Bq/cm$^3$.}
   \label{fig:outgas_vs_time}
\end{figure}

In this model, the time profile of removed tritium depends critically on the relative size of the de-trapping rate $\mu$ compared to the diffusion rate $D/a^2$. To illustrate this dependence, in Figure~\ref{fig:outgas_vs_time} we have plotted $M(t)$ for different ratios of $\mu$ and $D/a^2$, assuming an equal initial concentration of free and de-trappable tritium ($c_0 = s_0 = 1$ Bq/cm$^3$), with no additional trapped tritium. It can be seen that when the de-trapping rate is large compared to diffusion, the trapped tritium is released very quickly and the outgassed tritium follows a standard diffusion curve with all the tritium removed (purple line). When the de-trapping rate is very slow compared to diffusion (i.e. the trap energy becomes large compared to the thermal energy), the free interstitial tritium is removed following a standard diffusion curve with the de-trappable component only removed on much longer time scales (black line). Intermediate values give modified diffusion curves with exponential behavior at early or late times.

\subsection{Solutions using Green's Functions}
Alternatively, one can also solve the diffusion and de-trapping equations using Green's functions. The Green's function describing an infinite plate with zero concentration at the boundaries ($x=0$ and $x=2a$) is \cite{cole2011}

\begin{align}
    G_{X11}&(x,t;x_0,t_0) = \nonumber \\ 
    &\frac{1}{a}\sum_{m=1}^{\infty}\mathrm{e}^{\frac{-m^2\pi^2D}{4a^2}(t-t_0)}\sin\left(\frac{m\pi}{2a} x\right)\sin\left(\frac{m\pi}{2a} x_0\right)
    \label{eq:greensfunction}
\end{align}
This describes the time and spatial evolution of an injection of tritium at a point $x_0$ and time $t_0$ described by $\delta(x-x_0)\delta(t-t_0)$. The time evolution of a uniform concentration of tritium $c_0$ at time $t_0=0$ is the convolution of Equation~\ref{eq:greensfunction} over $x$:
\begin{align}
    c_f(x,t) &= c_0\int_0^{2a}G_{X11}(x, t;x_0,t_0=0)\mathrm{d}x_0 \nonumber \\
             &= \frac{c_0}{a}\sum_{m=1}^{\infty} \mathrm{e}^{\frac{-m^2\pi^2D}{4a^2}t}\sin\left(\frac{m\pi}{2a} x\right) \frac{2a}{m\pi}\big(1-\cos(m\pi)\big) \nonumber \\
             &= c_0\frac{4}{\pi}\sum_{n=0}^{\infty} \frac{\mathrm{e}^{-\frac{(2n+1)^2\pi^2D}{4a^2}t}\sin\left(\frac{(2n+1)\pi}{2a} x\right)}{2n+1} \nonumber \\
    c_f(x,t) &= c_0\frac{4}{\pi}\sum_{n=0}^{\infty} \frac{\mathrm{e}^{-k_n^2Dt}\sin(k_n x)}{2n+1}
\end{align}
where $k_n = (2n+1)\pi/2a$. This matches equation \ref{eq:simplediffusion} for the diffusion-only model with the boundaries shifted from $(-a,+a)$ to $(0,+2a)$. 

We can also use the Green's function to model the de-trapping tritium as a spatially uniform impulse injected over time with amplitude $\mu s = \mu s_0\mathrm{e}^{-\mu t}$ if we neglect re-trapping of free tritium. The evolution is the convolution of the injection function with $G$:
\begin{align}
c_t(x,t) &= s_0\mu\int_0^t\int_0^{2a} G_{X11}(x,t;x_0,t_0)\mathrm{e}^{-\mu t_0}\mathrm{d}x_0\mathrm{d}t_0 \nonumber \\
    &= s_0\mu\frac{4}{\pi}\int_0^t\sum_{n=0}^{\infty} \frac{\mathrm{e}^{-k_n^2D(t-t_0)}\sin(k_n x)}{2n+1}\mathrm{e}^{-\mu t_0}\mathrm{d}t_0 \nonumber \\
c_t(x,t) &=s_0 \mu \frac{4}{\pi}\sum_{n=0}^{\infty} \frac{\mathrm{e}^{-k_n^2Dt}-\mathrm{e}^{-\mu t}}{\mu-k_n^2D}\frac{\sin(k_nx)}{2n+1}
\end{align}
The total concentration is then the sum of the contributions from the initial free and initial de-trappable components: $c(x,t)=c_f + c_t$. 

The cumulative amount of tritium released, calculated from the time integral of the flux, is then
\begin{alignat}{2}
    M(t) = & 2D\int_0^t&&\frac{\partial c}{\partial x}(0,t_0)\mathrm{d}t_0 \nonumber \\
    =\,&\frac{4}{a}\sum_{n=0}^{\infty}&&\frac{1-\mathrm{e}^{-k_n^2Dt}}{k_n^2}c_0\ + 
    \label{eq:cuml_outgas_green1} \\
    & &&\left(\frac{1-\mathrm{e}^{-k_n^2Dt}}{k_n^2D} - \frac{1-\mathrm{e}^{-\mu t}}{\mu}\right)\frac{\mu D}{\mu-k_n^2D}s_0 \nonumber
\end{alignat}
And alternatively calculated as the total initial amount minus the free and de-trappable concentration remaining:
\begin{align}
    M(t) =\,& c_0L + s_0L\left(1-\mathrm{e}^{-\mu t}\right)-\int_0^{2a}c(x_0,t)\mathrm{d}x_0 \nonumber \\
    =\,& c_0L + s_0L\left(1-\mathrm{e}^{-\mu t}\right) \label{eq:cuml_outgas_green2} \\
    & - c_0\frac{4}{a}\sum_{n=0}^{\infty}\frac{\mathrm{e}^{-k_n^2Dt}}{k_n^2} \nonumber \\
    & - s_0\mu\frac{4}{a}\sum_{n=0}^{\infty}\frac{\mathrm{e}^{-k_n^2Dt}-\mathrm{e}^{-\mu t}}{k_n^2(\mu-k_n^2D)} \nonumber
\end{align}
These two expressions can be shown to be algebraically equivalent, but the latter form converges much more quickly when evaluated numerically. These expressions are also numerically equivalent to Equation~\ref{eq:cuml_outgas} when evaluated over sufficiently large $n$. 

\nocite{saldanha_2025_15653174}
\bibliography{references_cap}
\end{document}